\documentclass[reprint,
 superscriptaddress,
 amsmath,amssymb,
 aps, pra
]{revtex4-2}

	\usepackage[T1]{fontenc}
	\usepackage[utf8]{inputenc}
    \usepackage[pdftex,bookmarks=false]{hyperref}
    \usepackage{microtype}
    \usepackage{graphicx}
    \usepackage{dcolumn}
    \usepackage{bm}
    \usepackage{multirow}
    \usepackage{booktabs}
    \usepackage{xcolor}
    \usepackage{mathtools}
    \usepackage{mathptmx}
    \usepackage{upgreek}
    \usepackage{float}
    \usepackage{amsmath}
    \usepackage{amssymb}

    \hypersetup{
    	colorlinks = true, 
		citecolor = blue,
		bookmarksnumbered = true,
		urlcolor = blue
	}
	
    \DeclarePairedDelimiterX\braket[2]{\langle}{\rangle}{#1 \delimsize\vert #2}

\begin{document}
\title{Generation of a nodal line and Weyl points by magnetization reorientation in Co$_3$Sn$_2$S$_2$}
\date{\today}

\author{F.~Schilberth}
\thanks{These two authors contributed equally}
\affiliation{Experimentalphysik V, Center for Electronic Correlations and Magnetism, Institute for Physics, Augsburg University, D-86135 Augsburg, Germany} 
\affiliation{Department of Physics, Institute of Physics, Budapest University of Technology and Economics, M\H{u}egyetem rkp. 3., H-1111 Budapest, Hungary}
\email{felix.schilberth@physik.uni-augsburg.de}

\author{M.-C. Jiang}
\thanks{These two authors contributed equally}
\affiliation{Department of Physics and Center for Theoretical Physics, National Taiwan University, Taipei 10617, Taiwan}
\affiliation{RIKEN Center for Emergent Matter Science, 2-1 Hirosawa, Wako 351-0198, Japan}
\email{ming-chun.jiang@riken.jp}

\author{F. Le Mardelé}
\affiliation{Laboratoire National des Champs Magnétiques Intenses, LNCMI-EMFL, CNRS UPR3228, Univ. Grenoble Alpes, Univ. Toulouse, Univ. Toulouse 3, INSA-T, Grenoble and Toulouse, France}

\author{L. B. Papp}
\affiliation{Department of Physics, Institute of Physics, Budapest University of Technology and Economics, M\H{u}egyetem rkp. 3., H-1111 Budapest, Hungary}

\author{I. Mohelsky}
\affiliation{Laboratoire National des Champs Magnétiques Intenses, LNCMI-EMFL, CNRS UPR3228, Univ. Grenoble Alpes, Univ. Toulouse, Univ. Toulouse 3, INSA-T, Grenoble and Toulouse, France}

\author{M.~A.~Kassem}
\affiliation{Department of Materials Science and Engineering, Kyoto University, Kyoto 606-8501, Japan} 
\affiliation{Department of Physics, Faculty of Science, Assiut University, 71516 Assiut, Egypt} 

\author{Y.~Tabata}
\affiliation{Department of Materials Science and Engineering, Kyoto University, Kyoto 606-8501, Japan} 

\author{T.~Waki}
\affiliation{Department of Materials Science and Engineering, Kyoto University, Kyoto 606-8501, Japan} 

\author{H.~Nakamura}
\affiliation{Department of Materials Science and Engineering, Kyoto University, Kyoto 606-8501, Japan} 

\author{G.-Y. Guo}
\affiliation{Department of Physics and Center for Theoretical Physics, National Taiwan University, Taipei 10617, Taiwan}
\affiliation{Physics Division, National Center for Theoretical Sciences, Taipei 10617, Taiwan}

\author{M. Orlita}
\affiliation{Laboratoire National des Champs Magnétiques Intenses, LNCMI-EMFL, CNRS UPR3228, Univ. Grenoble Alpes, Univ. Toulouse, Univ. Toulouse 3, INSA-T, Grenoble and Toulouse, France}
\affiliation{Faculty of Mathematics and Physics, Charles University, Ke Karlovu 5, Prague, 121 16, Czech Republic}

\author{R. Arita}
\affiliation{RIKEN Center for Emergent Matter Science, 2-1 Hirosawa, Wako 351-0198, Japan}
\affiliation{Department of Physics, University of Tokyo, 7-3-1 Hongo Bunkyo-ku, Tokyo, 113-0033, Japan}

\author{I.~K\'ezsm\'arki}
\affiliation{Experimentalphysik V, Center for Electronic Correlations and Magnetism, Institute for Physics, Augsburg University, D-86135 Augsburg, Germany}

\author{S.~Bord\'acs}
\affiliation{Experimentalphysik V, Center for Electronic Correlations and Magnetism, Institute for Physics, Augsburg University, D-86135 Augsburg, Germany}
\affiliation{Department of Physics, Institute of Physics, Budapest University of Technology and Economics, M\H{u}egyetem rkp. 3., H-1111 Budapest, Hungary}
\affiliation{HUN-REN–BME Condensed Matter Physics Research Group, Budapest University of Technology and Economics, M\H{u}egyetem rkp. 3., H-1111 Budapest, Hungary}
\email{bordacs.sandor@ttk.bme.hu}

\begin{abstract}
Topological magnets exhibit fascinating physics like topologically protected surface states and anomalous transport. Although these states and phenomena are expected to strongly depend on the magnetic order, experimental evidence for their magnetic manipulation remains challenging. Here, we demonstrate the magnetic field control of the topological band structure in Co$_3$Sn$_2$S$_2$ by magneto-optical spectroscopy. We resolve a magnetic field-induced redshift of the nodal loop resonance as the magnetization is rotated into the kagome plane. Our material-specific theory, capturing the observed field-induced spectral reconstruction, reveals the emergence of a gapless nodal loop for one of the in-plane magnetization directions. The calculations show that the additionally created Weyl points for in-plane fields marginally contribute to the optical response. These findings demonstrate that breaking underlying crystal symmetries with external fields provides an efficient way to manipulate topological band features. Moreover, our results highlight the potential of low-energy magneto-optical spectroscopy in probing variations of quantum geometry.
\end{abstract}
\maketitle

\section{Introduction}
In recent years, intriguing topological states of matter were predicted and experimentally confirmed, among them topological insulators, Dirac and Weyl semimetals \cite{Chang2013, Liu2014, Lv2015}. As the interplay between magnetism and topology may fundamentally shape the band structure through symmetries dictated by the magnetic order, magnetic materials potentially allowing field control of the topological states are particularly appealing for applications in spintronics and quantum information technology \cite{Gilbert2021, Hasan2010}. While several theoretical studies indeed predict that magnetic fields can create or annihilate Weyl nodes and move their position in the band structure \cite{Burkov2011, Tabert2016, Fang2016, Ghimire2019}, experimental demonstrations of these effects are scarce. 

Magnetotransport experiments suggest the transformation of a nodal ring to Weyl points in EuP$_3$ \cite{Mayo2022} and in Co$_2$MnAl \cite{Li2020a}, whereas Shubnikov-de Haas oscillations propose a field-induced reconstruction of the Fermi surface in Co$_3$Sn$_2$S$_2$ \cite{Ye2022}. In addition, it was shown in MnSi that nodal planes can emerge only on $k$ planes parallel to an applied magnetic field \cite{Wilde2021}. While these examples prove that indeed the band structure in topological magnets can be highly sensitive to the magnetic state, they also highlight the difficulties of magnetotransport and magnetometry experiments to provide direct band-specific information. Low-energy spectroscopy may help to overcome these challenges, and may additionally provide direct information on the quantum geometry \cite{Ahn2020, Ahn2021,Onishi2024, Ghosh2024}. However, perhaps the only spectroscopic study on the magnetic control of band topology so far is the scanning tunneling spectroscopy (STS) work of Yin \textit{et al.}, implying a reconstruction of massive Dirac bands upon the rotation of a magnetic field in the kagome metal Fe$_3$Sn$_2$ \cite{Yin2018}. Due to the shortcomings of other spectroscopic methods in magnetic fields, here we use magneto-optical spectroscopy to reveal the magnetic field control of the band topology in Co$_3$Sn$_2$S$_2$ up to 34 T.

As a proof of this concept, here we investigate the evolution of the topological features of the band structure in Co$_3$Sn$_2$S$_2$ by measuring magneto-reflectance upon the gradual rotation of the magnetization from the $c$ axis to two perpendicular directions in the kagome plane. We observe a strong redshift of the previously reported giant magneto-optical resonance \cite{Schilberth2023} upon this magnetization reorientation. As only few bands lie around the Fermi energy, our material-specific theory can unambiguously reveal the origin of this shift: Due to the lower symmetry of the in-plane magnetized case, various Weyl nodes emerge by closing the spin-orbit coupling (SOC) induced gap at specific $k$ points or even recover a continuous gapless nodal line, depending on the orientation of the magnetization within the kagome plane. Our showcase study provides a deep insight into the mechanism of magnetically driven reconstructions of electronic band topology.

We choose the archetypical magnetic Weyl semimetal Co$_3$Sn$_2$S$_2$ due to its simple band structure close to the Fermi level. The crystal structure is built up by kagome layers of Co stacked along the $c$ axis in an $ABC$-fashion. Below $T_\text{C}\approx177$ K, it becomes a collinear ferromagnet with magnetic moments pointing along the stacking direction \cite{Zabel1979} with extraordinarily large uniaxial anisotropy, where fields as high as 23 T are required to fully turn the magnetization into the kagome plane \cite{Shen2019}. 
Nonrelativistic density functional theory (DFT) calculations predicted nodal loops, located around the Fermi energy on high symmetry planes of the Brillouin zone (BZ), which are gapped by SOC except for six Weyl points \cite{Wang2018}.
Angular resolved photoemission studies confirmed the existence of the nodal lines and traced their evolution across the magnetic transition \cite{Belopolski2021}. In the paramagnetic state, these results suggest the existence of a 4-fold degenerate Dirac loop, which upon breaking time-reversal by the magnetic order is split into two Weyl loops of opposite spin polarisation. One of them moves to higher energies upon increasing magnetic moment, while its partner stays close to the Fermi energy and was shown to be responsible for large anomalous Hall effect, Nernst effect and magneto-optical activity \cite{Belopolski2021, Wang2018, Ghimire2019, Minami2020, Okamura2020, Schilberth2023}. These results indicate that the topological band structure features are strongly coupled to the magnetic order in this kagome metal.

\section{Results}
\subsection{Symmetry analysis}
\begin{figure}
    \centering
    \includegraphics[width=\linewidth]{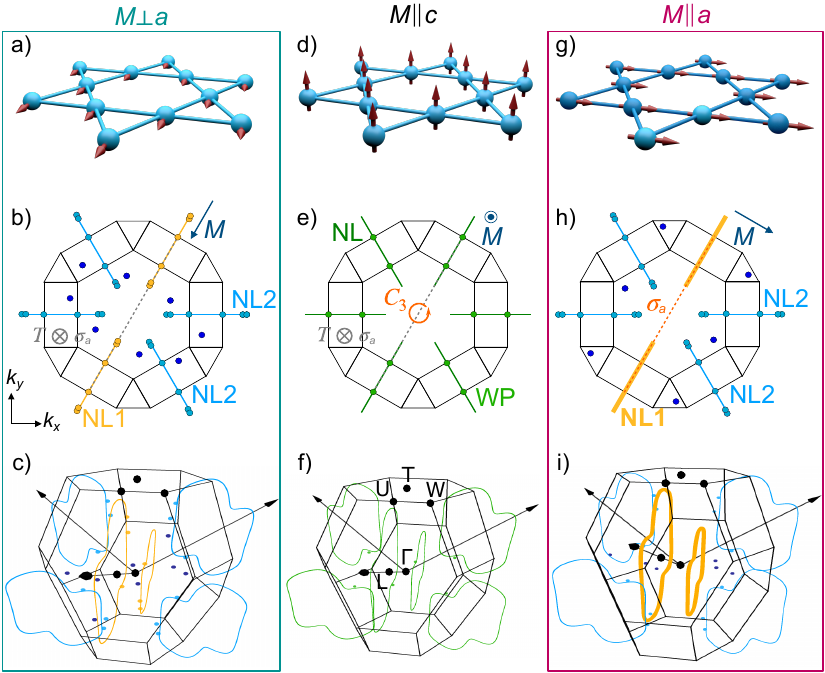}
    \caption{Symmetries and topology for out--of-- and in--plane magnetic field in Co\textsubscript{3}Sn\textsubscript{2}S\textsubscript{2}. The first row (a, d, g) shows a real--space representation of the magnetization for $\mathbf{M}\perp a$, $\mathbf{M}\parallel c$ and $\mathbf{M}\parallel a$. The second and third rows show a top and 3D view of the corresponding BZ for the respective configuration. (a--c) Inequivalent nodal loops (NL1/2) and emerging Weyl points (WP) for $\mathbf{M}\perp a$. Those have the same color as their corresponding loop or are dark blue if located at other $k$ points. (d--f) Three equivalent nodal loops related by the $C_{3}$ rotational symmetry for out--of--plane magnetization. (g--i) Mirror plane protected nodal loops for $\mathbf{M}\parallel a$.}
    \label{fig:Symmetries}
\end{figure}

In order to motivate our detailed experimental and \textit{ab initio} studies, we consider the modification of symmetries upon the field-induced magnetization reorientation. For the ground state magnetization $\mathbf{M}\parallel c$, all nodal loops are equivalent due to the $C_{3}$ rotational symmetry, as shown in Fig. \ref{fig:Symmetries}(d--f). In contrast, for magnetization in the kagome plane, the 3--fold rotational symmetry is broken, leaving mirror symmetries when the axial vector of the magnetic moment is perpendicular to the mirror plane. Therefore, for $\mathbf{M}\parallel a$, this results in the magnetic point group $C2/m$, which hosts a mirror plane $\sigma_a$. As shown in Fig. \ref{fig:Symmetries}(g--i), this plane contains a nodal loop NL1, which will therefore be protected from being gapped by SOC (thick yellow). The remaining loops, NL2 (light blue) are related by the mirror operation, but are not protected, hence they are gapped except for 6 Weyl points on each loop. Additionally, we find 6 Weyl points away from the nodal lines (dark blue). So in total we obtain 18 Weyl points and a protected nodal loop in the BZ. For the perpendicular in--plane direction $\mathbf{M}\perp a$, shown in panels (a--c), we obtain a similar situation. However, in this case, the magnetic point group is $C2^\prime/m^\prime$, e.g. the mirror plane is a symmetry only in combination with time-reversal. As this does not protect a pair of loops (NL1), these are gapped to 6 Weyl points by SOC. The remaining loops are again related by this operation and gapped to 6 Weyl points. Here we find 8 additional Weyl points away from the loops, so we obtain a total of 26 Weyl points distributed in the BZ. 

Some of these features were predicted earlier, e.g. Ghimire \textit{et al.} \cite{Ghimire2019} theoretically investigated the detailed evolution of the 26 Weyl points for $\mathbf{M}\parallel a$ upon canting of the magnetization, but did not comment on the gap closing of the fully degenerate loop. Similarly, Ozawa \textit{et al.} initially investigated the topological properties of a minimal tight-binding model for Co$_3$Sn$_2$S$_2$ \cite{Ozawa2019}, which was later extended to a more complete analysis of the magnetization direction dependent band structure including the fully degenerate nodal loop \cite{Ozawa2024}. While both groups comment on implications for the Hall conductivity, a direct experimental investigation of field-induced reconstructions of band topology has been lacking.

Along this line, the central question of this work is whether changes in the nodal lines and the emergence of additional Weyl nodes can be resolved by optical spectroscopy. Additionally, a very recent theoretical study by Nakazawa \textit{et al}. \cite{Nakazawa2024} developed for kagome monolayers, predicts an anisotropy of the band structure for different in-plane directions of the magnetization, which manifests in magnetotransport and, as we show, also in optical properties of bulk crystals.

\subsection{Magneto-optical spectroscopy}
\begin{figure*}
    \centering
    \includegraphics[width=\linewidth]{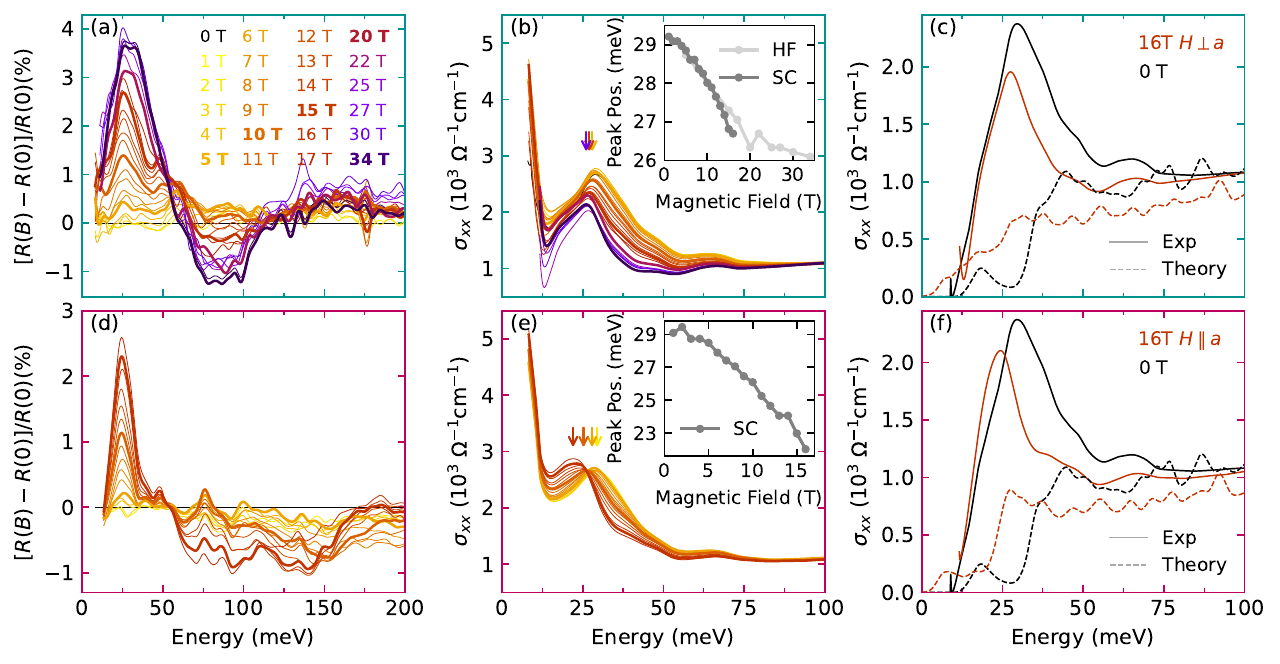}
    \caption{Magneto-reflectance data measured for fields up to 34 T. (a, d) Raw magneto-reflectance data $[R(B)-R(0)]/R(0)$ for $\mathbf{H}\perp a$ and $\mathbf{H}\parallel a$, respectively  (b, e) Optical conductivity in field calculated from the zero-field reflectivity of Ref.~\onlinecite{Schilberth2023} and the magneto-reflectance data for both in-plane field orientations. Insets: evolution of the 29 meV peak as a function of the field for the merged datasets from HF and SC experiments. The arrows, indicating the peak position for selected fields, highlight the stronger redshift for $\mathbf{M}\parallel a$  (5,15,34 T in (b) and 1,5,10,16 T in (e)). (c, f) Comparison of the zero-field and high-field conductivities with \textit{ab initio} calculations for both field directions.}
    \label{fig:Magnetoreflectance}
\end{figure*}

Fig.~\ref{fig:Magnetoreflectance}(a) and (d) show magneto-reflectance spectra over the energy range of 12-200 meV in different magnetic fields applied in the kagome plane along and perpendicular to the $a$ axis, respectively. Due to the limited measurement time at the high--field (HF) facility, the spectra for the latter configuration could only be collected in a superconducting (SC) setup up to 16 T. For both configurations, we obtain similar magneto-reflectance spectra, $[R(B)-R(0)]/R(0)$, although the detailed lineshapes are slightly different. With increasing magnetic field, a peak develops at low energies, followed by a minimum at the high-energy side. For $\mathbf{H}\perp a$, these two features are centered around 29 meV and 85 meV, while for $\mathbf{H}\parallel a$ the peak is shifted to lower energies and the minimum broadens considerably. For both configurations, the magneto-reflectance vanishes towards higher energies, indicating that field-induced band reconstructions are limited to the $\lesssim 200$ meV vicinity of the Fermi energy. 

The detailed field dependence of the low-energy peak for the two directions is shown in Fig. S1 for the SC and HF datasets, where error bars are determined from the noise level of the spectra. For $\mathbf{H}\perp a$ in panel (a), the magnitude increases up to 22 T and then saturates, following the same behavior as the magnetization for in--plane fields \cite{Shen2019}. The field dependence of the peak for $\mathbf{H}\parallel a$ is shown in panel (b), resulting in the same trend and similar magnitude up to 16 T. For $\mathbf{H}\perp a$, the additional spectral weight due to this peak pushes the reflectivity very close to 1 for high fields, as shown in Fig. S2. The observed changes with a peak magnitude of 3.7\% at 34 T can thus be considered remarkably large for an itinerant magnet. 

In order to quantify and interpret the magnetic field-induced changes of the band structure, we calculate the field dependence of the optical conductivity. The corresponding spectra are shown in Fig.~\ref{fig:Magnetoreflectance}(b) and (e) for the two field directions, exhibiting a very similar behavior in magnetic field. The most prominent features are the tail of the Drude response dominating the spectra below 15 meV and a peak around 29 meV, which shifts to smaller energies at higher fields. The position of the latter as a function of the field is shown in the respective inset. It again follows the same trend as the magnetization for both in-plane field directions, although the magnitude of the shift is larger for $\mathbf{H}\parallel a$.

As demonstrated in Ref.~\onlinecite{Schilberth2023}, this peak is caused by transitions between bands associated with the partially gapped nodal lines, which can be excited almost exclusively by one of the two circular polarization states of light. In panels (c) and (f), the spectra measured at 0 T (magnetization along $c$) and 16 T (magnetization almost in plane), after subtraction of the Drude contribution, are compared to the spectra obtained from \textit{ab initio} calculations for the two magnetic configurations. The theory spectra reproduce the experimental features on a broad energy scale (compare Ref. \onlinecite{Schilberth2023}). The shift of the nodal line resonance is captured well as the onset of optical weight appears at lower energies for both directions of the in-plane fields, though the peak in the theory spectra is not as pronounced as in the experiment. The difference may be caused by electronic correlations, or the magnitude of the matrix elements for these transitions may be underestimated by the theory. Especially the former is a likely scenario, as correlations would compress the band structure around the Fermi level. Hence, the energy difference in the denominator of the Kubo-Greenwood formula would strongly enhance the optical conductivity around resonances, specifically concentrating optical weight at low-energy interband excitations. Beyond this simple picture, Xu \textit{et al.} showed in their DFT+DMFT calculation, that indeed the optical conductivity at the 30 meV peak can be strongly enhanced by many-body effects \cite{Xu2020}. Due to the quantitative agreement of the experimentally observed and theoretically calculated redshifts, we nevertheless believe that our current DFT treatment is sufficient to explain the mechanism of the gap shrink.

We directly compare the calculated optical conductivity for different in-plane orientations of the magnetic field in Fig. S3. The in-plane spectra are very similar at high energies but show small differences around the resonance at 30 meV. While for $\textbf{M}\parallel a$ the slope of the peak is shifted parallel compared to the $\textbf{M}\parallel c$  spectrum, the peak for $\textbf{M}\perp a$ is broadened and the slope reduced. In addition, the redshift is larger for $\textbf{M}\parallel a$. All of these observations are consistent with the experimental data measured for the two directions.

\subsection{\textit{Ab initio} analysis}
\begin{figure}
    \centering
    \includegraphics[width=\linewidth]{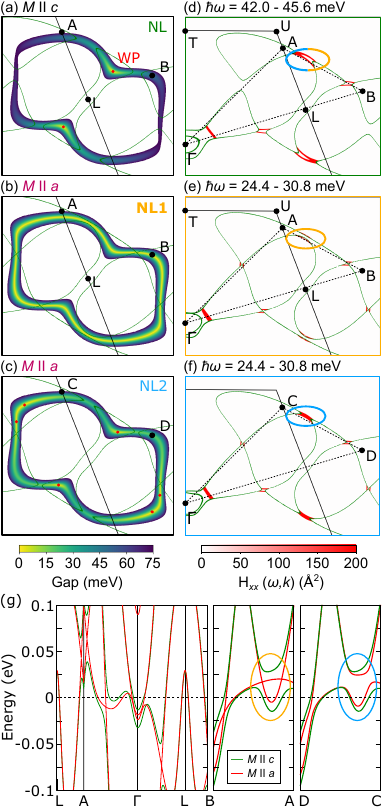}
    \caption{Evolution of the nodal loop upon reorienting the magnetization to $\textbf{M}\parallel a$. (a-c) gap of the nodal lines for out-of-plane and NL1 and NL2 in-plane loops, respectively. Points C and D are equivalent to A and B but lie on the BZ planes that contain NL2 instead of NL1. (d-f) Optical weight $H_{xx} (\omega,k)$ distribution at the peak energies (42.0-45.6 meV in (d), 24.4-30.8 meV in (e) and (f)) on the high symmetry planes containing the nodal loops from (a-c). (g) band structure along the triangles in the right column. The high symmetry points are shown above and in Fig. \ref{fig:Symmetries}(f). }
    \label{fig:Fig3}
\end{figure}

In order to evaluate the origin of the peak shift, we directly compare the band structures for the in- and out-of-plane configurations. Here, we only show the $\mathbf{M}\parallel a$ case with the protected nodal line, while the analysis for $\mathbf{M}\perp a$ can be found in Fig. S4, which yields very similar conclusions. Firstly, we analyze the differences between the nodal lines. Fig. \ref{fig:Fig3}(a-c) plot the gap size of the loops for $\mathbf{M}\parallel c$, NL1 and NL2 for $\mathbf{M}\parallel a$ respectively. The position of Weyl points is shown in red and the Fermi surface with dark green lines. For both in-plane loops in panels (b) and (c), the average gap size is smaller than for out-of-plane magnetization in panel (a), not only for the protected loop. This is clearly reflected in the band structure plot in panel (g), where we show the bands for $\mathbf{M}\parallel c$ and $\mathbf{M}\parallel a$ in green and red, respectively. Along A--B and C--D, the gapless nature of NL1 is highlighted, while the gap for NL2 is also smaller than for the bands with $\mathbf{M}\parallel c$.

Due to the smaller gap, these nodal lines contribute in a different energy range to the optical conductivity. Therefore, in panel (d), we plot the spectral weight distribution $H_{xx} (\omega,k$) (for a definition see Eq. \ref{eq:Hall-weight} in the Methods section) for out-of-plane magnetization at a photon energy of 42 - 45.6 meV, whereas for in-plane field in (e) and (f) we show $H_{xx}$ at 24.4 - 30.8 meV. Qualitatively, all plots show similar features with a small hotspot along $\Gamma$--L and several hotspots where the nodal lines lie around the Fermi energy, e.g. on the A--B and C--D lines. This comparison shows that the onset of optical weight of the optical conductivity for in- and out-of-plane magnetization is generated by the same band structure regions, but at different photon energies, explaining the redshift of the peak observed as the moments cant into the kagome plane. For the $\mathbf{M}\perp a$ case in Fig. S4, the spectral weight is plotted at energies between 28.8 - 33.6 meV, also explaining the redshift found for this configuration, although it is smaller than for $\mathbf{M}\parallel a$ which is also observed by the experiment.

Interestingly, if we infer the position of the Weyl nodes from the left column with the gap energy plots in Figs. \ref{fig:Fig3} and S4, we do not find any optical weight associated with transitions around these points for either magnetization direction. As was shown previously for $\mathbf{M}\parallel c$, the Weyl nodes are located around 60 meV above the Fermi level and therefore cannot contribute to the low-energy optical response \cite{Schilberth2023}. For in-plane fields, while the gap changes significantly, the energy of the crossing point with respect to the Fermi level does not change upon reorientation of the magnetization. Therefore, the same argument also accounts for the absence of optical weight from the Weyl points for in-plane field.

\section{Discussion}
First, let us discuss the possible origin of the nodal line reconstruction. Since the gap is initially produced by SOC, it has the same nature as the orbital magnetic moment~\cite{Guo1994} or the magnetocrystalline anisotropy~\cite{Wang1993,Jiang2022}. The origin of such anisotropy can be explained through the relative SOC strength by comparing the SOC matrix elements in the $p$-~\cite{Jiang2022} or the $d$-orbital basis~\cite{Takayama1976}. For the same spin channel, the SOC matrix elements $\langle d_{yz}\vert H_{\text{SOC}}\vert d_{xy,x^2-y^2}\rangle$ and $\langle d_{yz}\vert H_{\text{SOC}}\vert d_{z^2}\rangle$ prefer the in-plane anisotropy and $\langle d_{xy}\vert H_{\text{SOC}}\vert d_{x^2-y^2}\rangle$ and $\langle d_{yz}\vert H_{\text{SOC}}\vert d_{xz}\rangle$ prefer the out-of-plane anisotropy with the magnitude ratio being $\langle d_{yz}\vert H_{\text{SOC}}\vert d_{xy,x^2-y^2}\rangle^2:\langle d_{yz}\vert H_{\text{SOC}}\vert d_{z^2}\rangle^2:\langle d_{xy}\vert H_{\text{SOC}}\vert d_{x^2-y^2}\rangle^2:\langle d_{yz}\vert H_{\text{SOC}}\vert d_{xz}\rangle^2=1:3:4:1$~\cite{Takayama1976}. Fig. S5 shows the projected band structures of Co$_3$Sn$_2$S$_2$ without the SOC. We notice that the nodal lines consist mainly of Co $d_{xy,x^2-y2}$ orbitals with less contributions from the Co $d_{xz,yz}$ orbitals. Thus, from the above relation, we note that the SOC matrix element of $\langle d_{xy}\vert H_{\text{SOC}}\vert d_{x^2-y^2}\rangle$ dominates, which vanishes for in-plane spin orientation. This not only implies that out-of-plane magnetism is energetically preferred, but also that the SOC strength is weaker when the in-plane field is applied. Hence, we observe a shrinking of the gap for in-plane magnetization in Co$_3$Sn$_2$S$_2$. 

\begin{figure}
    \centering
    \includegraphics[width=\linewidth]{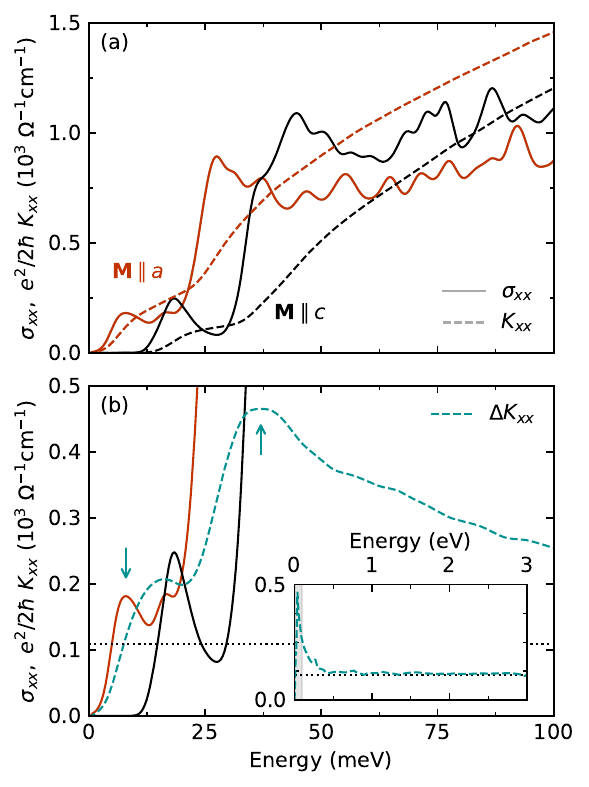}
    \caption{(a) Optical conductivity $\sigma_{xx}$ and the cutoff-frequency dependent quantum weight $K_{xx}$ of Co$_3$Sn$_2$S$_2$ under the in-plane ($\mathbf{M}\parallel c$) and out-of-plane ($\mathbf{M}\parallel a$) magnetic configurations. (b) Difference of the quantum weight $\Delta K_{xx}$ between in- and out-of-plane magnetization, which peaks exactly at the energy of the nodal line resonance, highlighting its critical origin by comparison with the optical conductivity (shown in the same line style as in panel (a)). The dotted horizontal line denotes the saturation value, visible more clearly in the inset which shows the enhancement on a broad energy range. The gray shading denotes the energy scale of the main panel.}
    \label{fig:Fig4}
\end{figure}

Next, we discuss the relation of our observation to the quantum geometry. The first moment of the absorptive optical conductivity $W^1=\operatorname{Re}(\sigma_{xx})/\omega$ is proportional to the quantum metric of the electronic states. The quantum geometry theory for the interband optical conductivity is developed considering a multiband manifold in the reciprocal space, leading to the definition of the multiband quantum metric, also known as the quantum geometric tensor, given as the product of the Berry connections \cite{Toermae2023, Ahn2021, Onishi2024}. Note that from Eq.~\ref{eq:Hall-weight}, the evaluation of the optical conductivity sums over all bands, which results in the quantum metric $g_{\alpha\beta}$ (known as the Fubini-Study metric~\cite{Ahn2021}). Importantly, through the delta function, we can probe the quantum metric for a certain transition energy. Accordingly, the quantum weight $K_{xx}$ has been introduced as an integrated measure of the quantum metric up to a specific cutoff frequency $\omega_c$. This relation has been formalized in recent works of Onishi et al.~\cite{Onishi2024}.
\begin{equation}
\label{eq: QGw}
\operatorname{Re} W_{x x}^1\left(\omega_c \rightarrow \infty\right) \equiv \int_0^{\omega_c \rightarrow \infty} d \omega \frac{\operatorname{Re} \sigma_{x x}}{\omega}=\frac{e^2}{2 \hbar} K_{xx}.
\end{equation}
In the following, we analyze the optical conductivity and quantum weights for in-plane, $\mathbf{M}\parallel a$ and out-of-plane, $\mathbf{M}\parallel c$ magnetization as displayed in Fig.~\ref{fig:Fig4}(a). We find that the redshift of the dominant optical transitions enhances the energy-resolved quantum metric $\operatorname{Re}(\sigma_{xx})/\omega$ through the denominator, subsequently enhancing the quantum metric, as evidenced by the larger quantum weight $K_{xx}$.

Quantum weight is always positive and would eventually saturate at a large enough cutoff frequency $\omega_c$. In Fig.~\ref{fig:Fig4}(b), we show the difference of quantum weight between the two magnetic configurations by defining $\Delta K_{xx}=\left(K_{xx}^{\mathbf{M}\parallel a}-K_{xx}^{\mathbf{M}\parallel c}\right)$.  On a larger energy scale shown in the inset, we can see that the quantum weight saturates at a cutoff frequency around 1 eV, with a difference in quantum weight being $(e^2/2\hbar)\Delta K_{xx}\sim$ 109 S/cm. Interestingly, we find that the saturated value of $\Delta K_{xx}$ is generated at the first optical resonance of $\mathbf{M}\parallel a$, indicated by the left arrow in Fig.~\ref{fig:Fig4}(b). Moreover, the largest quantum metric enhancement appears at the redshift of the band edge, or the spin-orbit resonance studied in this work (marked by the second arrow). Therefore, the critical quantum geometry enhancements by the magnetization reorientation appear at low frequencies, directly related to the nodal line reconstruction.

Through the Souza-Willkins-Martin sum rule~\cite{Souza2000}, the quantum weight is related to the electron localization length, which encodes the delocalization of the ground-state wavefunctions. That is, the stronger the quantum metric is, the stronger the wavefunctions overlap, the stronger the hopping amplitude becomes, the more metallic the states are, and the smaller the gap becomes. This explains why quantum geometry enhancement leads to a shrinking optical gap and redshift of the peak. We provide more details in Supplemental Note 1, including the applicability of the quantum weight theory to Co$_3$Sn$_2$S$_2$ (Eq.~\ref{eq: QGw}) as well as the comparison of hopping amplitudes near the band inversion of the nodal loop to support the observed quantum geometry enhancement in Co$_3$Sn$_2$S$_2$. Overall, we have demonstrated the experimental enhancement of quantum geometry in Weyl or Dirac semimetals with nodal crossings by low-frequency magneto-optical spectroscopy ~\cite{Ahn2020}.

In summary, we presented a detailed magneto-optical study to monitor the reconstruction of the topological band structure in a prominent kagome magnet upon manipulation of the magnetic order by external fields. We find that in Co$_3$Sn$_2$S$_2$, a low-energy resonance associated with a transition between bands of the nodal lines is highly sensitive to in-plane fields, namely it shows a red shift as the magnetization cants into the $ab$ plane. Our material-specific theory is able to reproduce this shift and shows that it originates from the reduction of the SOC-induced gap upon the rotation of the magnetization to the kagome plane. In particular, a complete degenerate nodal line emerges for $\mathbf{M}\parallel a$. Compared to quantum oscillation or STM measurements, this direct access to the interband response highlights the unique capabilities of optical spectroscopy to monitor changes in the band structure and its quantum geometry. As the dimension of the degenerate manifolds increases from Weyl points for $\mathbf{M}\perp a$ to loops for $\mathbf{M}\parallel a$, we expect an enhancement of the Hall response and reconstruction of the Hall spectral weight at finite frequencies. Following the changes in the topology of the bulk band structure, surface states may also be restructured. Since the nodal loop stays intact for $\mathbf{M}\parallel a$, we expect the formation of a drumhead surface state spanning the projection of NL1 on the surface, instead of the Fermi arcs present between the isolated Weyl points, with possible implications for e.g. surface catalysis of hydrogen evolution \cite{Rajamathi_2017, Jiang2022a}. The magnetization reorientation, therefore, provides an exciting pathway to study these exotic surface states, also applicable to other materials.

\section{Methods}

\subsection{Experimental Details}
We carried out magneto-reflectance experiments on the same $ab$ cut crystal which was used in a previous magneto-optical study \cite{Schilberth2023}. Relative magneto-reflectance spectra, $[R(B)-R(0)]/R(0)$ were collected in Voigt configuration with the field aligned perpendicular to the in-plane $a$ crystal axis, where $R(0)$ is the zero field and $R(B)$ is the finite field reflectivity spectrum. The light of a globar was analyzed by a Bruker Vertex 80v Fourier-transform spectrometer in the energy range between 12 and 500 meV and guided to the sample position by a light-pipe. The sample was cooled to 4 K and placed in a superconducting coil (SC) for measurements up to 16 T and in a resistive magnet for high field measurements (HF) up to 34 T. The reflected light was collected and guided onto an external bolometer after passing a silicon beamsplitter. Due to the different optical path lengths in the SC and HF experiments, the raw signal was smaller in the high-field data in the overlapping field range. In order to compare the two datasets, the HF data was scaled with the corresponding SC data set at 5 and 10 T, whereas for 15 T and higher fields, the 15 T SC data was used for scaling. Therefore, the presented data for high fields shows a lower limit for the actual magnitude of the magnetic field-induced changes. To calculate the optical conductivity in field, we multiply the smoothed magneto-reflectance spectra with the 10 K zero-field reflectivity spectrum from Ref.~\onlinecite{Schilberth2023} and perform a Kramers Kronig analysis with the same parameters as in the earlier work (see Fig.~S2).\vspace{0.5cm}

\subsection{DFT calculations}
The calculations of the electronic structures are conducted using DFT as implemented in the Vienna \textit{ab initio} simulation package (VASP)~\cite{Kresse1996a,Kresse1996b}. 
All the calculations are performed using the projector-augmented wave (PAW)~\cite{Blochl1994} pseudopotential with the generalized gradient approximation (GGA) in the form of Perdew-Burke-Ernzerhof (PBE)~\cite{Perdew1996,Kresse1999}. 
A plane-wave cutoff value of 400 eV and a $\Gamma$-centered $16\times 16 \times 16$ $k$-mesh is used to describe the electronic structure. 
The valence orbital set is $4s^23d^7$ for Co, $5s^25p^2$ for Sn, and $3s^23p^4$ for S, respectively. 
For the crystal structure, we use the experimental lattice parameters, $a$ = 5.379 \AA~and $\alpha$ = 59.8658$^\circ$~\cite{Vaqueiro2009}. 
Based on the DFT electronic band structures, we construct the Wannier functions using Co $d$, Sn $s$, $p$, and S $s$, $p$ orbitals. Finally, the optical properties are evaluated using the Kubo-Greenwood formula under the Wannier interpolation~\cite{Pizzi2020}. A fine mesh of 200 $\times$ 200 $\times$ 200 $k$ points is applied during the integration with good convergence. 

The optical conductivity given in Fig. 2(c) and the deduced spectral weight of certain photon energies plotted in Fig. 3 can be calculated through the Brillouin zone integration of the transition matrix elements:
 \begin{align}
	\label{eq:Hall-weight}
	\operatorname{Re}\sigma_{\alpha\beta}(\hbar\omega)&=\frac{-\pi\omega e^2}{\hbar}\int_\text{BZ}\sum_{nm}f^{\text{FD}}_{mn} A_{nm,\alpha}(\textbf{k})A_{mn,\beta}(\textbf{k})\delta(\omega_{mn}-\omega) \nonumber \\
	&=\frac{-\pi e^2}{\hbar}\int_\text{BZ}H_{\alpha\beta}(\hbar\omega;\textbf{k})\nonumber \\
        &=\frac{-\pi  e^2}{\hbar}\int_\text{BZ}\omega g_{\alpha\beta}(\hbar\omega;\textbf{k}) 
\end{align}
$\alpha$ and $\beta$ are the indices in Cartesian coordinates, $\int_{\text{BZ}}$ is the Brillouin zone integration $\int d^3\textbf{k}/(2\pi)^3$, $f^\text{FD}_n$ is the Fermi-Dirac distribution of band $n$, $f^\text{FD}_{mn} = f^\text{FD}_{m}-f^\text{FD}_{n}$. $\hbar\omega_{mn}=\hbar\omega_m-\hbar\omega_n$ is the energy difference between bands $m$ and $n$. Finally, $A_{nm,\alpha} = \left\langle u_{n,k} \middle | i \nabla_{k_\alpha} \middle |u_{m,k} \right\rangle$ is the Berry connection, $H_{\alpha\beta}(\hbar\omega)$ is the spectral weight at certain transition energy $\omega$, and $g_{\alpha\beta}(\hbar\omega)$ is the quantum metric~\cite{Toermae2023,Ahn2021}. They are all functions of the crystal momentum \textbf{k}. Due to the fixed proportionality between $H_{\alpha\beta}(\hbar\omega_{mn})$ and $g_{mn;\alpha\beta}(\hbar\omega)$ at a fixed photon energy, the maps in Figs. \ref{fig:Fig3}(d)-(f) and S4(d)-(f) can be equivalently interpreted as maps of the quantum metric.

\section{Declarations}

\subsection{Data availability}
The datasets generated during and/or analyzed during this study are available on Zenodo \href{https://doi.org/10.5281/zenodo.15354907}{10.5281/zenodo.15354907}.

\subsection{Acknowledgements}
This work was supported by the Hungarian National Research, Development and Innovation Office NKFIH Grants No. FK 135003 and by the Ministry of Innovation and Technology and the National Research, Development and Innovation Office within the Quantum Information National Laboratory of Hungary, by the National Science and Technology Council in Taiwan, by the IPA program of RIKEN, Japan and by the Deutsche Forschungsgemeinschaft (DFG, German Research Foundation) – TRR 360 – 492547816. We acknowledge the support of the LNCMI-CNRS, a member of the European Magnetic Field Laboratory (EMFL).

\subsection{Author Contributions}
M.A.K., Y.T., T.W. and H.N. synthesized and characterized the crystals; F.S. measured the zero-field reflectivity; F.S., F.L., L.B.P., I.M., M.O. and S.B. performed the magentoreflectance experiments. M.C.J., G.Y.G. and R.A. performed the \textit{ab initio} calculations; F.S., M.C.J. and S.B. wrote the paper; I.K. and S.B. planned and coordinated the project. All authors contributed to the discussion and interpretation of the experimental and theoretical results and to the completion of the paper.

Correspondence to Felix Schilberth or Sándor Bordács on the magnetoreflectance experiments or to Ming-Chun Jiang for the DFT calculations. 

\subsection{Competing interests}
The authors declare no competing interests.

\bibliography{References.bib}

\end{document}